\documentclass[letterpaper,aps]{revtex4-1}
\usepackage{graphics,amssymb,amsmath,epsfig,color}
\usepackage{graphicx}
\usepackage[english]{babel}

\makeatletter
\def\bbl@set@language#1{%
  \edef\languagename{%
    \ifnum\escapechar=\expandafter`\string#1\@empty
    \else\string#1\@empty\fi}%
  \@ifundefined{babel@language@alias@\languagename}{}{%
    \edef\languagename{\@nameuse{babel@language@alias@\languagename}}%
  }%
  \select@language{\languagename}%
  \expandafter\ifx\csname date\languagename\endcsname\relax\else
    \if@filesw
      \protected@write\@auxout{}{\string\select@language{\languagename}}%
      \bbl@for\bbl@tempa\BabelContentsFiles{%
        \addtocontents{\bbl@tempa}{\xstring\select@language{\languagename}}}%
      \bbl@usehooks{write}{}%
    \fi
  \fi}
\newcommand{\DeclareLanguageAlias}[2]{%
  \global\@namedef{babel@language@alias@#1}{#2}%
}
\makeatother

\DeclareLanguageAlias{en}{english}
\DeclareLanguageAlias{EN}{english}

\begin{document}

\title{Thermo-Optic Multi-Stability and Relaxation in Silicon Microring Resonators with Lateral Diodes}

\author{Dodd Gray}
 \email{dodd@mit.edu.}

\author{Ryan Hamerly}%
 \altaffiliation[Also at ]{NTT Research Inc., East Palo Alto, CA 94303, USA}
\affiliation{ 
Research Laboratory of Electronics, Massachusetts Institute of Technology, 77 Massachusetts Avenue, Cambridge, MA 02139, USA
}%

\author{Meysam Namdari}
\altaffiliation[Now at ]{Fraunhofer Institute for Photonic Microsystems, Maria-Reiche-Str. 2, 01109 Dresden, Germany}
\affiliation{Integrated Photonic Devices Laboratory, Technische Universit\"at Dresden, Helmholtzstr. 16, 01069 Dresden, Germany}

\author{Mircea-Traian C\u atuneanu}
\affiliation{Integrated Photonic Devices Laboratory, Technische Universit\"at Dresden, Helmholtzstr. 16, 01069 Dresden, Germany}

\author{Nate Bogdanowicz}
 \altaffiliation[Now at ]{Entanglement Technologies Inc., 42 Adrian Court, Burlingame, CA 94010, USA}

\author{Hideo Mabuchi}
\affiliation{Ginzton Laboratory, Stanford University, 348 Via Pueblo Mall, Stanford, California 94305, USA}

\author{Kambiz Jamshidi}
\affiliation{Integrated Photonic Devices Laboratory, Technische Universit\"at Dresden, Helmholtzstr. 16, 01069 Dresden, Germany}




\email{\authormark{*}dodd@mit.edu} 

\date{\today}
\begin{abstract}
We demonstrate voltage-tunable thermo-optic bi- and tri-stability in silicon photonic microring resonators with lateral p-i-n junctions and present a technique for characterizing the thermo-optic transient response of integrated optical resonators. Our method for thermo-optic transient response measurement is applicable to any integrated photonics platform and uses standard equipment. Thermo-optic relaxation in encapsulated waveguides is found to be approximately logarithmic in time, consistent with the analytic solution for 2D heat diffusion. We develop a model for thermo-optic microring multi-stability and dynamics which agrees with experiment data over a wide range of operating conditions. Our work highlights the fundamental connection in semiconductor waveguides between active free-carrier removal and thermo-optic heating, a result of particular relevance to Kerr soliton state stability and on-chip frequency comb generation. The devices studied here were fabricated in a CMOS foundry process and as a result our model is useful for design of silicon photonic waveguide devices.
\end{abstract}
\maketitle

\section{Introduction}

The microring resonator is a basic building block in integrated photonic systems, providing spectral filtering and resonant interaction enhancement for myriad applications in optical sensing \cite{trocha_ultrafast_2018}, communication \cite{pfeifle_coherent_2014,sun_single-chip_2015} and information processing \cite{okawachi_coupled_2019}. Thermo-optic (TO) tuning and dynamics in microrings must often be considered in the design of integrated optical systems, especially those employing large circulating optical powers or requiring high precision \cite{carmon_dynamical_2004,johnson_self-induced_2006,atabaki_sub-100-nanosecond_2013,de_cea_power_2019}. A noteworthy example is the generation of optical frequency combs from continuous-wave pump lasers, where the ratio of thermal relaxation and soliton formation timescales has been shown to determine whether frequency combs can be stably formed in microrings \cite{li_stably_2017}. Experimental characterization of microring thermal relaxation is thus critical for integrated frequency comb device design and control \cite{joshi_thermally_2016}.

Thermo-optic dynamics are prominent in silicon due to its high TO coefficient and the presence of two-photon absorption (TPA), which leads to carrier generation and heating \cite{johnson_self-induced_2006}.  Furthermore, to reduce optical loss, extraction of mobile free carriers from waveguides with a reverse-biased p-i-n junction is often employed.  While carrier extraction can reduce the lifetime of carriers by nearly two orders of magnitude \cite{turner-foster_ultrashort_2010,gajda_design_2011,sen_effect_2013}, providing access to new dispersive nonlinear optical regimes \cite{yu_mode-locked_2016,hamerly_conditions_2018}, the bias voltage provides a multiplier effect that enhances the TO effect.  Therefore, accurate models and characterization techniques for TO dynamics are especially important in silicon photonics.

Here we present experimental characterization of TO dynamics and steady state behavior in silicon microring resonators with lateral p-i-n junctions for FC extraction fabricated in a CMOS foundry process. We present a dynamical model for TO nonlinearities in silicon waveguides,  compare the fixed points of the model with experimental data, and show that they are in good agreement. As part of this study we demonstrate a novel method for measuring thermal relaxation in semiconductor resonators. This technique uses two tunable CW lasers tuned near distant longitudinal resonances of the same microring to pump and probe decaying TO detuning after TO bistability jumps. This relatively simple and low-cost technique makes use of standard equipment and provides a direct measurement of TO detuning decay with high resolution in time and waveguide temperature. We show that in encapsulated silicon microring resonators TO detuning is approximately logarithmic in time, consistent with two-dimensional heat flow from the waveguide into the surrounding encapsulant \cite{jaeger_conduction_1959}.

\begin{figure}[ht!]
\centering\includegraphics[width=6.45in]{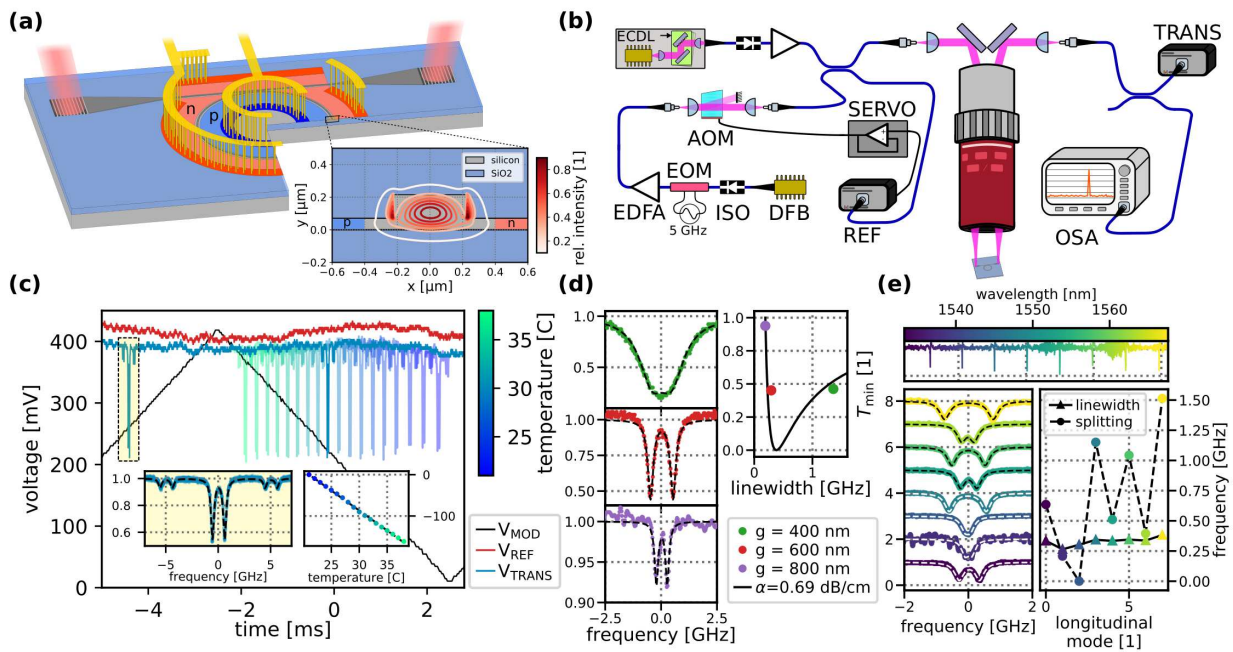}
\caption{ (a) Schematic depiction of silicon photonic microring resonator with lateral p-i-n diode for free carrier extraction. Inset shows the optical mode intensity profile overlaid on the waveguide cross-section. (b) Experiment schematic, discussed in the text. (c) Bus waveguide transmission traces measured at sample temperatures between 15--30~C. Left inset shows zoomed transmission dip with doublet profile and sidebands for tuning rate calibration. Right inset shows linear thermal tuning with a tuning coefficient $\delta_{\rm TO}\approx-9.67\,\mathrm{GHz}\,\mathrm{K}^{\rm -1}$. (d) Resonance profiles of identical resonators with varied coupler gaps. (e) Fine structure of eight consecutive longitudinal modes of the same microring, showing large variation in splitting.}
\label{fig:f1}
\end{figure}

\section{Device and Experiment}\label{sec_experiment}

Here we study thermal relaxation in a silicon lateral-junction microring resonator produced by a multi-project wafer service at a CMOS foundry (IMEC, Belgium) \cite{absil_imec_2015}. The microring device was fabricated in the top silicon layer of a silicon-on-insulator (SOI) substrate directly above a 2~$\mu$m buried oxide (BOX) layer of the SOI substrate and is encapsulated from above by amorphous silica with approximately 10~$\mu$m thickness. The device geometry and waveguide mode profile are illustrated in Fig.~\ref{fig:f1}(a). The optical resonator is formed by a 40~$\mu$m-diameter ring of silicon ridge waveguide with $215\times450\,\mathrm{nm}$ cross section and 70~nm slab height (inset to Fig.~\ref{fig:f1}(a)). The microring is evanescently coupled to a straight bus waveguide terminated on both ends with grating couplers. A radial p-i-n junction diode is formed across the waveguide by p-doping (n-doping) the the 70~nm silicon cladding inside (outside) the ring; this facilitates carrier extraction for tunable control of the carrier lifetime.


Fig.~\ref{fig:f1}(b) depicts the experimental setup used to characterize this device. The chip was attached to a ceramic chip-carrier with silver paint and wire bonded to provide electrical access to the lateral p-i-n diode. The chip-carrier was placed on a temperature-controlled mount attached to a 3-axis translation stage for positioning under a microscope objective for free-space coupling. Two diode laser sources---an external cavity diode laser (ECDL, tuning range 1530--1570~nm) and a distributed feedback laser (DFB) operating near 1550~nm with a 200~GHz ($\approx$1.6~nm) mode-hop-free current tuning range---were coupled into the input port after optical isolation with Faraday isolators (ISO) and amplification with Erbium-doped fiber amplifiers (EDFAs). The DFB laser power was actively stabilized to enable frequency modulation via current injection at fixed optical power. This was accomplished using a reference detector (REF) and a servo circuit (SERVO) controlling an acousto-optic modulator (AOM). The DFB was also passed through a 20~GHz electro-optic amplitude modulator (EOM) to enable addition of known-frequency sidebands for frequency sweep calibration. Light collected from the output grating was fiber coupled, power split and simultaneously sent to a transmission photoreceiver (TRANS) and an optical spectrum analyzer (OSA) for detection. Optical power signals measured by the REF and TRANS photorecievers were recorded using an oscilloscope. The OSA and oscilloscope, as well as input light conditioning, sample temperature and reverse bias voltage were all computer-controlled with \textsc{Python} code using the \textsc{Instrumental} software package \cite{bogdanowicz_instrumental_2018}.





The linear optical resonance properties of the microring were initially probed using only the DFB laser with approximately 1~$\mu$W optical power coupled into the bus waveguide.  With a fixed reverse-bias voltage $V_{\rm rb}$ applied across the ring p-i-n diode, transmission spectra are collected by sweeping the DFB frequency over a 180~GHz range using a 30~Hz triangle-wave current modulation.  Temperature-dependent bus waveguide transmission data are shown in Fig.~\ref{fig:f1}(c). The left inset of Fig.~\ref{fig:f1}(c) shows a single resonance transmission dip with Lorentzian doublet lineshape and 5~GHz sidebands, and the right inset shows tuning of the resonance center frequency with sample temperature. Individual resonances were found to have a loaded $Q$ (linewidth) of 5.4$\times$10$^5$ (360~MHz) with a TO tuning coefficient $\delta_{\rm TO}\approx-9.67\,\mathrm{GHz}\,\mathrm{K}^{\rm -1}$. Spectroscopy of otherwise-identical resonators with varied directional coupler gaps, shown in Fig.~\ref{fig:f1}(d), revealed the intrinsic Q (linewidth) to be $\approx$8.8$\times$10$^5$ (220~MHz), corresponding to 0.7~dB/cm waveguide loss in the ring. Lineshapes of several consecutive longitudinal resonances of the same device measured using the ECDL are shown in Fig.~\ref{fig:f1}(e). The Lorentzian doublet splitting was found to vary randomly between ~0--1.5~GHz from one longitudinal mode to the next on a single device, indicative of roughness-induced standing-wave resonances \cite{weiss_splitting_1995,hosseini_high_2009}.

\begin{figure}[ht!]
\centering\includegraphics[width=5.3in]{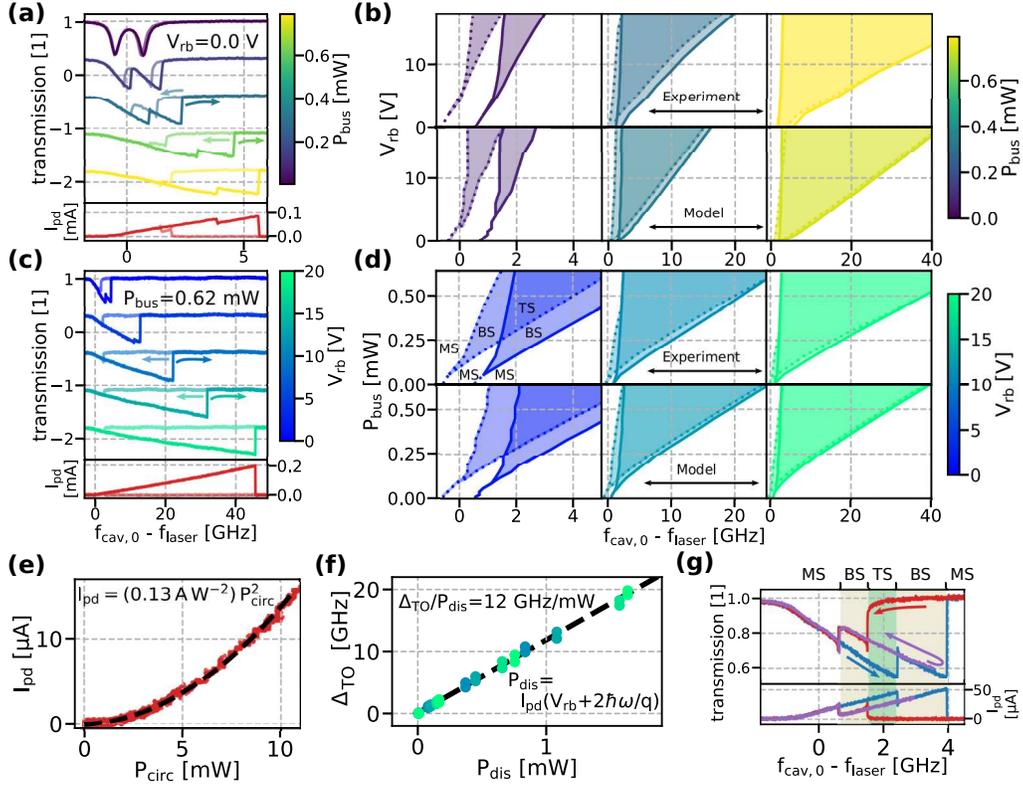}
\caption{(a) Power- and (b) voltage-dependent transmission through the bus waveguide, showing voltage-tunable TO bistability. Darker (lighter) curves represent blue-to-red (red-to-blue) laser tuning, as indicated by the arrows. The red curves at the bottom are lateral diode current corresponding to bottom transmission curve. (c) Experimentally measured (top and third rows) and calculated (second and fourth rows) regions of TO multistability as a function of reverse bias voltage (top two rows) and input power (bottom two rows). Mono-, bi- and tri-stability are labeled as MS, BS and TS respectively. (d) Measured correspondence between circulating optical power and lateral diode photocurrent with a quadratic fit. (e) Measured TO tuning and voltage-controlled power dissipation with linear fit. (f) Measured optical transmission and lateral diode current showing TO tri-stability. Multistable detuning ranges are labeled as (c).}
\label{fig:f2}
\end{figure}

\section{Tunable Thermo-Optic Multistability}\label{sec_multi}

As the optical input power was increased, voltage ($V_{\mathrm{rb}}$)-dependent asymmetric and hysteretic distortion of the resonance lineshape due to TO nonlinearity became observable. At $V_{\mathrm{rb}}=0\,\mathrm{V}$ and $15\,\mathrm{V}$ the onsets of noticeable TO distortion were $\geq$~80~$\mu$W and $\geq$~10~$\mu$W in the bus waveguide, respectively. Measured TO resonance distortion and bistability for varied optical power and reverse bias voltage are shown in Fig.~\ref{fig:f2}(a)-(b). Lateral diode photocurrent $I_{\mathrm{pd}}$ was synchronously collected and analyzed. Fig.~\ref{fig:f2}(d) shows $I_{\mathrm{pd}}$ to vary quadratically with circulating power $P_{\mathrm{circ}}$. Despite recent reports of dominantly linear-absoprtion mediated free carrier generation at 1550~nm up to 	$>$10~mW power levels in similar silicon waveguides \cite{gil-molina_optical_2018,dainese_generation_2019}, we did not observe significant linear photocurrent generation down to $\mu$W levels. Assuming high quantum efficiency of photocurrent collection, $V_{\mathrm{rb}}$ and $I_{\mathrm{pd}}$ also enable calculation of the TO heat load $P_{\mathrm{dis}}=I_{\mathrm{pd}}(V_{\mathrm{rb}} + 2\hbar\omega/q )$, where $\hbar\omega\approx0.8\,\mathrm{eV}$ is the photon energy and $q$ is the electron charge. On-resonant TO detuning $\Delta_{\mathrm{TO}}$ and $P_{\mathrm{dis}}$ measured for input optical powers 10--800$\mu$W and reverse bias voltages 0--20~V are plotted in Fig.~\ref{fig:f2}(e) along with a linear fit. Note that the vast majority of resonator-coupled optical power is not included in $P_{\mathrm{dis}}$ because it is lost to scatter and does not induce local heating. Interestingly in this device we observe TO tri-stability in the regime where $\Delta_{\mathrm{TO}}$ exceeds the roughness-induced normal-mode splitting, indicated by the presence of three distinct jump positions in the forward and backward frequency sweep data. This is highlighted in Fig.~\ref{fig:f2}(f), which shows transmission and $I_{\mathrm{pd}}$ traces when the laser frequency is swept (at fixed power and $V_{\mathrm{rb}}$) across the resonance in both direction and when the sweep is reversed before the bistable jump.


To understand the voltage-tuned TO multistabilities in this device we use a previously presented model for coupled nonlinearities in microrings \cite{hamerly_conditions_2018,hamerly_optical_2017}, modified to include two field amplitudes for orthogonal standing-wave modes and with parameters inferred from the linear and TO characterization data shown in Figures~\ref{fig:f1} and \ref{fig:f2}. Measured and calculated phase diagrams showing voltage-tunable regions of TO mono-, bi- and tri-stability as functions of bus-waveguide-coupled power $P_{\rm bus}$, cold cavity detuning $\Delta$ and reverse-bias voltage $V_{\rm rb}$ are shown in Fig.~\ref{fig:f2}(c). The model details and the parameter values used are included in Appendix~\ref{sec:model}.



\begin{figure}[ht!]
\centering\includegraphics[width=5.3in]{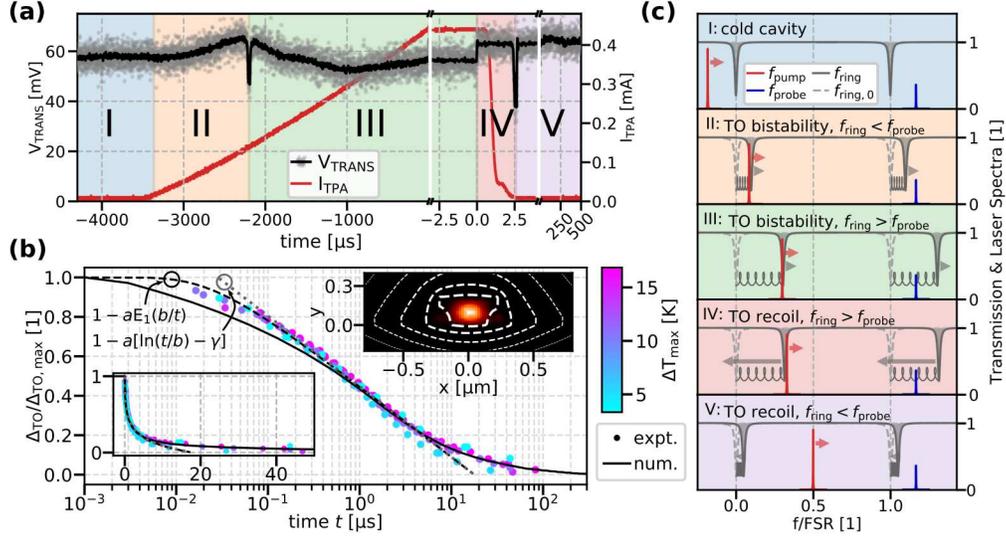}
\caption{Direct measurement of microring thermal relaxation with two CW lasers. (a) Example trace showing two transmission dips and one small jump in measured with a single photodiode as a strong pump laser is tuned across a microring resonance, inducing TO bistability. The dips result from a longitudinal mode passing a fixed-wavelength probe laser before and after the TO bistability jump. The delay after the jump as a function of probe laser detuning is used to measure the time profile of thermal relaxation. (b) Schematic spectra showing the evolution of the cavity, pump and probe laser spectra through the five labeled stages in (a). (c) Measured, numerically modeled, and fit analytical time profiles of microring TO relaxation, normalized to initial temperature deflection. Thermal relaxation is approximately logarithmic in time for 100~ns--10~$\mu$s time delays, as discussed in text. Relaxation from temperature deflections from ambient between 4--16~K were measured, as indicated by the color bar on the right. Insets show the same data plotted on a linear timescale and the initial temperature distribution around the optical mode profile. }
\label{fig:f3}
\end{figure}

\section{Thermo-Optic Relaxation Measurement}\label{sec_to_relaxation}

The bistable jumps seen at high optical powers and reverse bias voltages enable direct measurement of transient TO relaxation by monitoring the detuning of a different longitudinal mode of the microring with a second, weak ``probe'' laser. Here the 1550~nm DFB and ECDL sources discussed above were used as the ``pump'' and the ``probe'', respectively. A schematic representation, divided into five stages, of the dynamic TO tuning of the microring's longitudinal modes relative to the pump and probe lasers is shown in Fig.~\ref{fig:f2}(a). The ECDL was tuned in steps near a longitudinal mode of the ring at $\approx$1530~nm while the DFB frequency was scanned, driving repeatable TO bistable jumps. Traces collected for a single value of $\Delta f_{\mathrm{probe}}$ exemplifying this measurement scheme are shown in Fig.~2(b). A 1530~nm optical bandpass filter was used to block pump light at the transmission detector. The timing of the first and second probe resonance dips during a single scan indicate, respectively, the cold-cavity detuning of the probe $\Delta f_{\mathrm{probe}}$ and the time delay after the jump before the transient TO detuning falls back to $\Delta f_{\mathrm{probe}}$. A distinct jump also appears in each transmission trace due to a small fraction of pump light leaking through the bandpass filter. This provides an accurate reference for $t=0$ (the jump time) within a single trace. The synchronously recorded transverse diode current (also shown in Fig.~\ref{fig:f3}(a)) is consistent with this interpretation of the transmission data.

Relaxation of the microring waveguide temperature deflection $\Delta T (t)$, inferred from the probe resonance detuning, was measured over nearly four orders of magnitude in time delay after the jump (between 20~ns and 80~$\mu$s) with a temperature resolution of 0.1~K using different cold-cavity detuning values of the probe laser. The pump power was varied to control $\Delta T_\mathrm{max}$, the waveguide temperature deflection at time of the jump. Measurements were performed with $V_{\mathrm{rb}}=15\,\mathrm{V}$ to simultaneously maximize the achievable waveguide temperature deflection and minimize the impact of free carriers. Relaxation of the microring waveguide temperature after a TO bistability jump was observed to decay with an approximately logarithmic time dependence, in agreement with 2D numerical simulations of TO tuning from TPA-proportional heating in these waveguides as well as previous studies of transient heating in optical tweezers \cite{celliers_measurement_2000}

\begin{eqnarray}
    \Delta T(t) &\approx&   1 - a \mathrm{E_1}(b/t) \label{eq:E1_empirical} \\
                &\approx&         1 - a \left[\ln(t/b) - \gamma \right] \label{eq:log_empirical}
\end{eqnarray}


\noindent where $\mathrm{E_1}(x)=\int_x^\infty \frac{e^{-x}}{x}dx$ is the exponential integral function, $t$ is the time after the bistability jump, $\gamma\approx0.577$ is the Euler-Mascheroni constant and $a=0.155$ and $b=16\,\mathrm{ns}$ are constants determined by fitting to experimental data in Fig.~\ref{fig:f3}(b) encoding the heat capacity and thermal conductivity of the media surrounding the waveguide. $\mathrm{E_1}(x)$ has been approximated with a truncated series expansion in Eq.~\ref{eq:log_empirical}.  This TO relaxation time dependence is expected for 2D radial heat transfer from a point source (the waveguide cross section) into the distributed heat capacity of the surrounding medium \cite{jaeger_conduction_1959}, as discussed in Appendix~\ref{sec:log_decay}.

\section{Conclusions}\label{sec_conclusions}

We have measured and modeled TO effects in CMOS silicon microring resonators with lateral p-i-n junctions for active free carrier removal. Our results highlight that active free carrier removal comes with the side-effect of increased TO nonlinearity caused by Joule heating and demonstrate a voltage-tunable TO response that is potentially useful for a variety of waveguide devices requiring thermal control. We have mapped out measured regions of TO multistability in our device and specified a model for TO dynamics which shows good agreement and which can be used for future photonic device design in similar platforms. 

The demonstrated voltage-tunable increase in TO response can be viewed as a resource, providing a means of low power, high bandwidth electronic control over in-situ laser heating of the optical mode volume in a microring resonator. TO tuning of photonic resonators with ``auxiliary'' laser light has been shown to provide robust access to single-soliton states in microring Kerr combs \cite{geng_kerr_2017,zhang_sub-milliwatt-level_2019} and has been explored in a variety of optical media \cite{tenenbaum_all-optical_2018,hill_all-optical_2020}. While temperature control with guided laser light offers advantages in response time and power efficiency over conventional displaced electronic heaters, it requires large optical powers and is generally incompatible with semiconductor microrings where free carriers generated by two-photon absorption (TPA) of the auxiliary laser would add significant loss for all resonator modes. The reverse-bias regime explored here simultaneously reduces laser power required for TO tuning to the microwatt scale and the free carrier lifetime to ~10~ps, rendering nonlinear FC losses insignificant. This suggests a hybrid electro-optic approach may be viable for semiconductor waveguides which directly heats the optical mode volume without added loss and is voltage-tunable.

We have demonstrated a novel method of measuring the time profile of thermal relaxation in microring resontators using an auxiliary CW laser as a probe of dynamic detuning after TO bistability jumps. To our knowledge this is the first demonstration of this relatively simple and low-cost measurement technique, which is applicable for general characterization of TO dynamics in integrated optical waveguides. We have shown TO relaxation of silicon microrings fabricated in a standard CMOS process to be approximately logarithmic in time, a result of the quasi-2D nature of radial heat transfer from the waveguide into the surrounding materials, with the majority of cooling occuring in the first 1-2~$\mu$s and a long tail extending to nearly 80~$\mu$s. This measurement also highlights the potential for CMOS microrings as optically addressable thermometers with high temporal and temperature resolution, potentially useful for monitoring thermal dynamics in proximate microscopic (e.g.~biological) systems. Fabrication in a CMOS foundry process also provides for scalable arrays of sensors addressable via a single optical fiber with integrated detection and electronic signal processing.

\section*{Acknowledgments}
This work is supported in part by the German Research Foundation (DFG JA 2401/11) and by the National Science Foundation (PHY-1648807).


\appendix

\section{Cavity Multistability Model}\label{sec:model}

The thermo-optic (TO) multi-stability of the silicon microring resonator was modeled using the following set of coupled ordinary differential equations (ODEs), adapted from recent work studying free carrier oscillations (FCO) and Kerr comb generation \cite{hamerly_conditions_2018,hamerly_optical_2017}. In this model linearly orthogonal standing-wave modes with complex amplitudes $a_{\rm s}$ and $a_{\rm c}$ (for ``sine'' and ``cosine'') couple via two-photon abosorption (TPA) to a free-carrier population with density $n$ and temperature deflection from ambient $T$. The set of coupled ODEs is written in terms of unitless field, free carrier density and temperature quantities $\bar{a}_{\rm i} = a_{\rm i}/\xi_{\rm a}$, $\bar{n} = n/\xi_{\rm n}$ and $\bar{T} = T/\xi_{\rm T}$ normalized reflect the strengths of the Kerr, FCA and TO nonlinear optical coupling, respectively \cite{hamerly_conditions_2018,hamerly_optical_2017}:

\begin{eqnarray}
    \frac{d\bar{a}_{\rm c}}{d\bar{\tau}} &=& \left[ \left(-\frac{1}{2} + i\left[ \bar{\Delta}_0  - \frac{\bar{\delta}_{\rm r}}{2} \right]\right) +\left(i - r\right)\bigl(|\bar{a}_{\rm c}|^2 + \tfrac{2}{3}|\bar{a}_{\rm s}|^2\bigr) + \left(-i - \mu^{-1}\right)\bar{n}_{\rm c} + i\bar{T} \right]\bar{a}_{\rm c} + \sqrt{\eta}\bar{a}_{\rm in} \label{a_c_dynamics}\\
    \frac{d\bar{a}_{\rm s}}{d\bar{\tau}} &=& \left[ \left(-\frac{1}{2} + i\left[ \bar{\Delta}_0  + \frac{\bar{\delta}_{\rm r}}{2} \right]\right) +\left(i - r\right)\bigl(|\bar{a}_{\rm s}|^2 + \tfrac{2}{3}|\bar{a}_{\rm c}|^2\bigr) + \left(-i - \mu^{-1}\right)\bar{n}_{\rm s} + i\bar{T} \right]\bar{a}_{\rm s} + \sqrt{\eta}\bar{a}_{\rm in} \label{a_s_dynamics}\\
    \frac{d\bar{n}}{d\bar{\tau}} &=& -\frac{\bar{n}}{\bar{\tau}_{\rm fc}} +  \chi\left( |\bar{a}_{\rm c}|^4 + \frac{4}{3}|\bar{a}_{\rm c}|^2|\bar{a}_{\rm s}|^2 + |\bar{a}_{\rm s}|^4 \right)  \label{n_dynamics}\\
    \frac{d\bar{T}}{d\bar{\tau}} &=& -\frac{\bar{T}}{\bar{\tau}_{\rm th}} +  \zeta\left[ \phi \chi  \left(|\bar{a}_{\rm c}|^4 + \frac{4}{3}|\bar{a}_{\rm c}|^2|\bar{a}_{\rm s}|^2 + |\bar{a}_{\rm s}|^4\right) + \frac{ \chi \bar{n} }{\mu r} \left( |\bar{a}_{\rm c}|^2 + |\bar{a}_{\rm s}|^2 \right)  \right]\label{T_dynamics}
\end{eqnarray}

\noindent where the timescale is normalized to the photon lifetime ($\bar{\tau}=t/\tau_{\rm ph}$), $\bar{\Delta}_0 \pm \bar{\delta}_{\rm r}/2  = (\omega_{\rm laser} - \omega_{\rm cav})\tau_{\rm ph}$ are the cold-cavity detunings measured in resonance linewidths and $\bar{\delta}_{\rm r}=2\pi\delta_{\rm r}\tau_{\rm ph}$ is the linewidth-normalized normal-mode splitting due to backscattering in the waveguide ring resonator. Unitless parameters $\chi$ and $\zeta$ are normalized coefficients for FCD due to TPA and TO coupling coefficients, while $r$ and $\mu$ account for the Kerr/TPA and FCD/FCA nonlinearity strength ratios, respectively. Free carriers and temperature deflections decay on timescales $\bar{\tau}_{\rm fc}$ and $\bar{\tau}_{\rm th}$, respectively. The free carrier lifetime $\tau_{\rm fc}$ and dissipation-per-absorption parameter $\phi$ depend on $V_{\rm rb}$ as

\begin{eqnarray}
\phi &=& \frac{qV_{\rm rb} + 2\hbar\omega}{2\hbar\omega} \label{eq:phi_def}\\
\tau_{\rm fc} &\approx& \left(\tau_{\rm fc,0}-\tau_{\rm fc,sat}\right)e^{-V_{\rm rb}/V_{\tau_{\rm fc}}} + \tau_{\rm fc,sat}  \label{eq:tau_fc}
\end{eqnarray}


\noindent where $\hbar\omega$ is the photon energy ($0.8\,\mathrm{eV}$ at 1550$\,$nm), $\tau_{\rm fc,0}$ is the free carrier lifetime for $V_{\rm rb}=0$ (short-circuit condition), $\tau_{\rm fc,sat}$ is the minimal free-carrier lifetime limited by velocity saturation and $V_{\tau_{\rm fc}}$ is an empirical decay constant of $\tau_{\rm fc}$ with $V_{\rm rb}$. Assuming efficient free carrier collection from the undoped waveguide, the lateral photodiode current is proportional to the electron-hole pair generation rate

\begin{eqnarray}
I_{\rm pd} &=& \xi_{\rm I} \chi \left(| \bar{a}_{\rm c}|^4 + \frac{4}{3}|\bar{a}_{\rm c}|^2|\bar{a}_{\rm s}|^2 + |\bar{a}_{\rm s}|^4 \right) \label{eq:i_pd}\\
\xi_{\rm I} &=&  \frac{q v_{\rm m} \xi_{\rm n}}{2 \tau_{\rm ph}} \label{eq:xi_i}
\end{eqnarray}

\noindent where $v_{\rm m}=\pi d_{\rm ring} A_{\rm eff}$ is the mode volume and unity quantum efficiency is assumed. For our device parameters the current normalization is $\xi_{\rm I} \approx 5.9\,\mathrm{\mu A}$. When the normal mode splitting exceeds the resonance linewidth we expect two-photon absorption by the individual standing-wave modes to dominate the cross-coupling, such that $\left(| \bar{a}_{\rm c}|^4 + |\bar{a}_{\rm s}|^4 \right) \gg \frac{4}{3}|\bar{a}_{\rm c}|^2|\bar{a}_{\rm s}|^2$ in Equations~\ref{n_dynamics} and \ref{eq:i_pd}. Because the device geometry, resonance linewidth and material parameters in $\xi_{\rm n}$ are well determined, Equation~\ref{eq:i_pd} provides a robust measurement the intracavity optical power independent of alignment-sensitive grating coupler efficiencies which we observed to drift slowly over time. Thus $I_{\rm pd}$ was used to calibrate the instantaneous intra-cavity and bus-coupled optical powers $P_{\rm circ}$ and $P_{\rm bus}=\eta P_{\rm circ}/2$ reported in the main text, where the factor of two in the preceding relation appears because were excited uni-directionally but couple equally to forward- and backward-propagating waves.

Even at moderate reverse bias voltages $V_{\rm rb}>1$~V in our device we expect $\tau_{\rm fc} \ll \tau_{\rm ph}$, rendering FCA and FCD insignificant compared to the enhanced TO response. In this limit the TO shift $\Delta_{\rm TO}=\xi_{\rm T}\bar{T}\delta_{\rm TO}$ is proportional to the dissipated power $P_{\rm dis}= I_{\rm pd} \left( V_{\rm rb} + 2\hbar\omega/q \right)$. From these definitions and Eq.~\ref{T_dynamics}--\ref{eq:xi_i} we find that in steady state

\begin{eqnarray}
    \Delta_{\rm TO}   &\approx& \xi_{\rm T}  \left( \bar{\tau}_{\rm th} \zeta \right) \frac{r\phi}{\chi} \frac{ I_{\rm pd} }{\xi_{\rm I} }  \\
                        &=& \left( \bar{\tau}_{\rm th} \zeta \right) \frac{\xi_{\rm T}}{\xi_{\rm I}} \frac{\delta_{\rm TO} }{2\hbar\omega/q} P_{\rm dis} \label{eq:D_TO_vs_P_dis}
\end{eqnarray}

As expected $\Delta_{\rm TO}$ and $P_{\rm dis}$ were measured to be proportional experiment (see Fig.~\ref{fig:f2} (e) in the main text) with a ratio $\Delta_{\rm TO}/P_{\rm dis}\approx12$~GHz/mW. According to Eq.~\ref{eq:D_TO_vs_P_dis} we infer that $\bar{\tau}_{\rm th} \zeta \approx 16.3$. 

With the exception of $\tau_{\rm fc,0}$, all device-dependent parameters needed for steady-state solutions of these equations were determined from the experimental data presented in Figures 1 and 2 of the main text. A finite-element model (\textsc{Lumerical Device}) was used to estimate $\tau_{\rm fc,0}$ and $\tau_{\rm fc}(V_{\rm rb})$ with an empirical exponential fit, consistent with previous experimental and numerical studies of similar waveguides \cite{turner-foster_ultrashort_2010,gajda_design_2011}. Parameters values and normalization coefficients $\xi_{\rm i}$ are provided in Table~\ref{tab:t0}. Citations for bulk silicon material parameters are provided in previous work \cite{hamerly_conditions_2018}. 

The steady-state solutions of Eq.~\ref{a_c_dynamics}--\ref{T_dynamics} (along with the complex conjugates of Eq.~\ref{a_c_dynamics}--\ref{a_s_dynamics}) reduce to a simultaneous pair of seventh-order polynomial equations in $\left|\bar{a}_{\rm c}\right|^2$ and $\left|\bar{a}_{\rm s}\right|^2$, which was solved numerically with \textsc{Mathematica} in the experimentally accessible parameter range. Stability of the fixed points was determined from the locations in the complex plane of ODE system's Jacobian eigenvalues. The white, light and dark shaded regions in Fig.~\ref{fig:f2}(c) of the main text, respectively labeled MS, BS and TS, indicate parameter regimes in which one, two and three stable solutions were found. 

\begin{table}
\caption{\label{tab:t0} Normalization coefficients, definitions and relevant device and bulk silicon material parameters needed for TO multistability model given in Eq.~\ref{a_c_dynamics}--\ref{T_dynamics}. Device parameters are taken from experimental data as discussed in the text. Citations for bulk silicon material parameters are provided in previous work \cite{hamerly_conditions_2018}.}
\begin{ruledtabular}
\begin{tabular}{lrlll}

\multicolumn{1}{l}{Parameter} & \multicolumn{1}{l}{Symbol} & \multicolumn{1}{l}{} & \multicolumn{1}{l}{Definition}                                                         & Value               \\ 
\multicolumn{5}{c}{Normalization constants}                                                                                                                                                                                            \\ \hline
circulating field                        & $\xi_{\rm a}$                       & $=$                            & $\left[ \gamma v_{\rm g} \tau_{\rm ph}\right]^{-1/2}$                                            & 7913 $^{1/2}$ cm$^{-1}$     \\
input/ouput field                        & $\xi_{\rm in}$                      & $=$                            & $\left[ \gamma v_{\rm g} \tau_{\rm ph}^2  t_{\rm R}^{-1} \right]^{-1/2}$                  & 478 W$^{1/2}$ cm$^{-1}$      \\
free carrier density                     & $\xi_{\rm n}$                       & $=$                            & $2 / \left(\mu \sigma v_{\rm g} \tau_{\rm ph} \right)$                                           & $1.3\times10^{15}$ cm$^{-3}$  \\
temperature deflection                   & $\xi_{\rm T}$                       & $=$                            & $\left[\delta_{\rm TO} v_{\rm g} \tau_{\rm ph} \right]^{-1}$                                     & 36 mK          \\
diode current                           & $\xi_{\rm I}$                       & $=$                             & $ \frac{\pi d_{\rm ring} A_{\rm eff}}{2} \frac{ q \xi_{\rm n} }{ \tau_{\rm ph}}$                  & 5.9 $\mu$A                      \\ \hline
\multicolumn{5}{c}{Normalized coefficients}                                                                                                                                                                                            \\ \hline
TPA/SPM strength ratio                   & $r$                                 & $=$                            & $\frac{\beta}{2\gamma}$                                                                          & 0.19                          \\
FCD/FCA strength ratio                   & $\mu$                               & $=$                            & $\frac{4\pi}{\lambda}\frac{\left|dn_{\rm si}/dn_{\rm c}\right|}{d\alpha_{\rm si}/dn_{\rm c}}$    & 30                           \\
TPA FC generation coeff.                 & $\chi$                              & $=$                            & $\left( r \mu \sigma \right) / \left( 2 \hbar \omega \gamma v_{\rm g} \right)$                   & 13.8                          \\
TO tuning coeff.               & $\zeta$                             & $=$                            &                                                                                                 & 0.47                        \\ \hline
\multicolumn{5}{c}{Device and material parameters}                                                                                                                                                                                     \\ \hline
photon lifetime                          & $\tau_{\rm ph}$                     & $=$                            & $2\pi$/FWHM                                                                                      & 455 ps                       \\
max FC lifetime                          & $\tau_{\rm fc,0}$                   & $=$                            &                                                                                                  & 140 ps                       \\
min. FC lifetime                         & $\tau_{\rm fc,sat}$                  & $=$                           &                                                                                                  & 3.1 ps                        \\
FC lifetime decay const.                & $V_{\tau_{\rm fc}}$                   & $=$                           &                                                                                                  & 2.15 V                        \\
TO decay time                            & $\tau_{\rm th}$                     & $=$                            &                                                                                                  & 30 ns                        \\
round-trip time                          & $t_{\rm R}$                         & $=$                            & $\frac{\pi d_{\rm ring}}{v_{\rm g}} = 1/\mathrm{FSR}$                                            & 1.7 ps                       \\
normal mode splitting                    & $\delta_{\rm r}$                    & $=$                            &                                                                                                  & 1.1 GHz                      \\
TO tuning coefficient                    & $\delta_{\rm TO}$                   & $=$                            &                                                                                                  & -9.7 GHz K$^{-1}$            \\
ring-bus power coupling                  & $\theta$                            & $=$                            &                                                                                                  & $1.6\times10^{-3}$            \\
mode effective area                      & $A_{\rm eff}$                       & $=$                            &                                                                                                  & $1.0\times10^{-9}$ cm$^{2}$   \\
modal group index                        & $n_{\rm g,eff}$                     & $=$                            & $c / v_{\rm g}$                                                                                  & 3.97                         \\
ring diameter                            & $d_{\rm ring}$                      & $=$                            &                                                                                                  & 40 $\mu$m                    \\
photon energy                            & $\hbar\omega$                       & $=$                            &                                                                                                  & 0.8 eV                       \\
SPM coefficient                          & $\gamma$                            & $=$                            & $\frac{2\pi}{\lambda} n_{\rm 2}$                                                                 & $3.1\times10^{-9}$ cm/W       \\
FCA cross section                        & $\sigma$                            & $=$                            &                                                                                                  & $1.45\times10^{-17}$ cm$^{2}$  \\ 
\end{tabular}
\end{ruledtabular}
\end{table}


%


\section{Logarithmic Thermo-optic Relaxation}\label{sec:log_decay}

A simplified model can be used to understand the time dependence of the thermal transients in this system. We consider the step function
transient response of the 2D heat equation to a point heat source at the origin:

\begin{eqnarray}
\left[\nabla^2 - \frac{1}{\alpha} \frac{\partial}{\partial t} \right]\Delta T(r,t) &=& -\frac{Q(r,t)}{\kappa} \\
Q(r,t) &=& Q_0 \delta(r) \Theta(t) \label{eq:heat_source}
\end{eqnarray}

\noindent where $\alpha=\kappa/(\rho c_{\rm p})$ is the thermal diffusivity, $\kappa$ is the
thermal conductivity, $\rho$ is the density, $c_{\rm p}$ is the heat capacity, $Q_0$ is the linear power density of the heat source, $\delta(r)$ is the Dirac $\delta$-function at the origin and
$\Theta(t)$ is the Heaviside step function indicating that the heat source
is turned on at time $t=0$. For concreteness, we note that $Q_0=P_{\rm dis}/(\pi d_{\rm ring})$ in the context of microring heating by guided light. The solution for the temperature deflection transient
$\Delta T(r,t)$ can be expressed as an exponential integral function $\mathrm{E_1}$

\begin{eqnarray}
\Delta T(r,t) &=& \frac{Q_0}{4\alpha}\int_{\frac{r^2}{4\alpha t}}^\infty \frac{e^{-z}}{z}dz\\
&=& \frac{Q_0}{4\alpha} \mathrm{E_1}\left(\frac{r^2}{4\alpha t}\right)  \\
&\stackrel{t \gg r^2/4\alpha}{\longrightarrow}& \frac{Q_0}{4\alpha} \left[-\gamma - \ln\left( \frac{r^2}{4\alpha t}\right)\right] \label{eq:E1_puiseux}
\end{eqnarray}

\noindent where $\mathrm{E_1}(x)=\int_x^\infty \frac{e^{-x}}{x}dx$ is the exponential integral function, $\gamma\approx0.577$ is the Euler-Mascheroni constant and in Eq.~\ref{eq:E1_puiseux} $\mathrm{E_1}\left[r^2/(4\alpha t)\right]$ has been approximated by a truncated Puiseux series along the real axis \cite{sloane_handbook_2014}, valid in the limit $t \gg r^2/4\alpha$. We consider an approximately Gaussian transverse mode of an optical waveguide centered at the origin with waist $w_0$. $\Delta T(r,t)$ is sampled by waveguide modes to determine the transient TO perturbation. The point-source approximation made in Eq.~\ref{eq:heat_source} is self consistent at times $t \gtrsim w_0^2 /(4\alpha)$ when $\Delta T(r,t)$ is approximately constant over the mode intensity profile. From Eq.~\ref{eq:E1_puiseux} we expect that in this regime the transient temperature deflection overlapping the optical mode can be approximated as

\begin{eqnarray}
    \Delta T(t) &\approx&   1 - a \mathrm{E_1}(b/t) \label{eq:E1_empirical2} \\
                &\approx&         1 - a\left[\ln(t/b) - \gamma \right] \label{eq:log_empirical2}
\end{eqnarray}

\noindent where $a=0.155$ and $b=16\,\mathrm{ns}$ are constants determined by fitting to experimental data in Fig.~\ref{fig:f3}(b) that depend on the waveguide geometry and the heat capacity, thermal conductivity and refractive indices of the core and cladding materials. In a microring with diameter $d_{\rm ring}$ this two-dimensional heat transfer model will remain valid for times $t \lesssim d_{\rm ring}^2/(16\alpha)$. From the modal effective area $A=0.1\,\mathrm{\mu m}^2$ and the thermal diffusivity of amorphous silica $\alpha_{\rm a-SiO_2}=9\times10^{-3}\,\mathrm{cm}^2/\mathrm{s}$ we can estimate the transient TO step response to have the form of Eq.~\ref{eq:log_empirical} for $10\,\mathrm{ns}\lesssim t \lesssim 100\,\mathrm{\mu s}$. This rough estimate provides a reasonable prediction of the experimentally observed TO step response shown in Fig.~\ref{fig:f3}(b), however it neglects heat conduction by the 60~nm-thick crystalline silicon cladding in the plane of the microring and numerous other features of the device geometry. We used a finite element model (\texttt{Lumerical DEVICE}) to account for these aspects and still found a logarithmic transient consistent with Eq.~\ref{eq:log_empirical}. Our numerical model suggests that the high thermal diffusivity of the thin silicon cladding layer ($\alpha_{\rm si}=0.8\,\mathrm{cm}^2/\mathrm{s}$) can account for the experimentally observed deviation from logarithmic time dependence near $t=10\,\mu\mathrm{s}$ (Fig.~\ref{fig:f3}(b)).




%

\end{document}